\documentclass{aa}
\usepackage{natbib,psfig}
\begin{document}

% \thesaurus{02.07.1; 03.13.8; 11.03.4;
% 11.05.1; 11.05.2; 11.06.1; 11.06.2; 11.19.6}

\title{An XMM-Newton view of the extended ``filament'' near the
cluster of galaxies Abell~85}

\author{F. Durret\inst{1} \and G.B. Lima Neto\inst{2} \and 
W. Forman\inst{3} \and E. Churazov\inst{4}}

\institute{
Institut d'Astrophysique de Paris, CNRS, 98bis Bd Arago, 75014 Paris, France 
\and
Instituto de Astronomia, Geof\'{\i}sica e C. Atmosf./USP, R. do Mat\~ao 1226, 05508-090
S\~ao Paulo/SP, Brazil 
\and
Harvard Smithsonian Center for Astrophysics, 60 Garden St, Cambridge MA
02138, USA
\and
MPI f\"ur Astrophysik, Karl Schwarzschild Strasse 1, 85740 Garching, Germany
}

\date{Accepted ??/2003. Received ??/2003; Draft printed: \today}

\authorrunning{Durret et al.}
\titlerunning{The X-ray filament near Abell~85}

\abstract{ We have observed with XMM-Newton the extended 4~Mpc
filament detected by the ROSAT PSPC in the neighbourhood of the
cluster of galaxies Abell~85.  We confirm that there is an extended
feature, aligned at the same position angle as the major axis of the
central cD, the bright cluster galaxies, and nearby groups and
clusters.  We find that the X-ray emission from the filament is best
described by thermal emission with a temperature of $\sim2$ keV, which
is significantly lower than the ambient cluster medium, but is
significantly higher than anticipated for a gas in a weakly bound
extended filament. It is not clear whether this is a filament of
diffuse emission, a chain of several groups of galaxies, or stripped gas
from the infalling south blob.  }

\maketitle

\section{Introduction}\label{sec:intro}

The complex of clusters Abell~85/87/89 is a well studied structure,
whose dominant component is Abell~85 at a redshift of $z=0.056$
\citep{Durret98op}.\footnote{At $z=0.056$, 1 arcmin $= 90
h_{50}^{-1}\,$kpc, assuming $\Omega_M=1$, $\Omega_\Lambda=0$.}
Abell~89 comprises two background structures and will not be discussed
further (see \cite{Durret98X} for details).

In a combined X-ray (ROSAT PSPC) and optical (imaging and large
redshift catalogue) analysis \citep{Durret98X}, we have shown the
existence of an elongated X-ray structure to the southeast of the main
cluster Abell~85. This filament is at least 4~Mpc long (projected
extent on the sky), and it is not yet clear whether this is a filament
of diffuse emission or a chain made by several groups of galaxies.

The search for filaments or strings of groups lying along filaments
visible in X-rays is obviously an exciting topic, since such
structures could trace the filaments of dark matter along which large
scale structures form, as commonly seen in N-body simulations of
structure formation in cold dark matter (CDM) cosmological models
\citep[e.g.][]{Frenk83,Jenkins98}. Large scale filaments are also
observed in the distribution of galaxies in redshift surveys
\cite[e.g.][]{Schmalzing} and between clusters \citep{West95}. A large
fraction of the baryons in the Universe are predicted to be in diffuse
form in the network of filaments traced by the CDM. Although the
detection of X-ray emitting gas in filaments has not always been
successful \citep{Briel,Pierre}, a few X-ray filamentary structures
have been detected in Coma \citep{Vikhlinin}, Abell~2125 \citep{Wang},
and between Abell~3391 and Abell~3395 \citep{Tittley}.

Based on ROSAT data, \cite{Durret98X} showed that after subtracting
the modelled contribution of Abell~85 the remaining emission has the
appearance of being made of several groups, each of these small
structures having an X-ray luminosity of about $10^{42}$ erg~s$^{-1}$,
typical of small groups of galaxies.  In order to analyse better the
morphology and physical properties of the X-ray gas in the filament,
we have reobserved Abell~85 with XMM-Newton, and present our first
results below. With the greater sensitivity of XMM-Newton, we are able
to estimate the temperature and metal abundance of the X-ray emitting
gas in the inner part of the filament; however, we are still unable to
decide whether the gas is diffuse or concentrated in group like
structures.  Note that Kempner et al. (2002) discuss the Chandra
observation of Abell 85 with emphasis on the X-ray concentration at
the northernmost end of the filamentary structure, but the filament
itself lies outside their field.

\section{Observations and data reduction}\label{sec:obs}

Two XMM-Newton \citep{Jansen01} observations were performed on January
7th, 2002. The first exposure was pointed between the centre of
Abell~85 (coincident with the cD galaxy) and the southern substructure
(the ``South Blob''); the second observation was aimed at the southern
filament.  Both total exposure times were 12.5~ks using the
\textsc{medium} optical filter in standard Full Frame mode. The basic
data processing (the ``pipeline'' removal of bad pixels, electronic
noise and correction for charge transfer losses) was done with SAS
V5.3.

For the spectral analysis, we have used the southern filament observation made
with both EPIC/MOS cameras. After applying the standard filtering, keeping
only events with FLAG=0 and PATTERN $\le$ 12, we have removed the observation
times with flares using the light-curve of the [10--12 keV] band. The cleaned
MOS1 and MOS2 event files have remaining exposure times of 12.36~ks and
12.40~ks, respectively.

The redistribution and ancillary files (RMF and ARF) were created with the SAS
tasks \texttt{rmfgen} and \texttt{arfgen} for each camera, taking into account
the extended nature of the filament.

The smoothed, merged image of the four cleaned MOS observations in the
0.3--5.0 keV energy band is shown in Fig.~\ref{fig:rawimage}. To
generate the image, we subtracted background from each data set (taken
from Lumb et al.  2002), generated exposure maps, smoothed the
background subtracted data and the exposure maps, and finally
corrected for the exposure. Besides the well known main cluster and
South Blob, emission can clearly be seen extending southeast of the
South Blob. This confirms our detection of the ``filament'' with the
ROSAT PSPC. Note from Fig.~\ref{fig:rawimage} that Abell~87 is not
detected in X-rays by XMM-Newton, as it was already not detected by
the ROSAT PSPC \citep{Durret98X} or by the ROSAT All Sky Survey (which
did detect the south blob). This strongly suggests that this is not
really a cluster but perhaps a concentration of several groups, in
agreement with Katgert et al. (1996) who find two different redshifts
in that direction, one coinciding with that of Abell 85 and the other
at z=0.077.

\begin{figure}[!htb]
\centering
%\mbox{\psfig{figure=durret.fig1.ps,width=8.5cm,height=8.5cm}}
\caption[]{Cleaned, background subtracted, exposure map corrected
image obtained from the two MOS1 and MOS2 XMM-Newton exposures. The
image has been smoothed with a Gaussian of $\sigma=5\,$arcsec.  }
\label{fig:rawimage}
\end{figure}

In order to analyse the structure of the filament in more detail, we
constructed a model of the overall cluster X-ray emission from an azimuthally
averaged profile and subtracted it from the merged image, as previously done
with the ROSAT data. The result is displayed in Fig.~\ref{fig:imminusmodel}.
The brightest structure coincides with the South Blob and the filament is
clearly seen to be patchy, as already observed with ROSAT (see Fig.~2 in
\cite{Durret98X}). The positions of the various patches of X-ray emission do
not exactly coincide with those previously detected with ROSAT, but the
overall structure is comparable.

\begin{figure}[!htb]
%\centering \mbox{\psfig{figure=durret.fig2.ps,width=8.5cm}}
\caption[]{Merged XMM-Newton image obtained after subtracting an
azimuthal average of the X-ray emission of the overall cluster. The
filament is clearly seen. The ellipse (E1) covering the filament shows the
region where we extracted the events for the spectral analysis (the
three circles were excluded). The straight line dividing this ellipse
in two defines the ``north'' and ``south'' halves of the filament. 
The ellipse to the left (E2) indicates the region where the cluster
contribution was estimated. }
\label{fig:imminusmodel}
\end{figure}

\section{Physical properties of the filament}

In order to analyse in more detail the physical properties of the
filament and derive its temperature and metallicity, we extracted
events inside an elongated region E1 of elliptical shape centered at
R.A. $=0^{\rm h}42^{\rm m}15.0^{\rm s}$ and Decl. $=
-09^{\circ}39^{\prime}24^{\prime\prime}$ (J2000), with major and minor
axes of 8 and 4~arcmin, respectively; three circular regions
corresponding to point sources were excluded (see
Fig.~\ref{fig:imminusmodel}). We also extracted events in a second
elliptical region (E2) having the same size as E1, with its major axis
parallel to that of E1 and displaced towards the east. E2 is taken to
be representative of the diffuse cluster emission.

The background was taken into account by extracting spectra (for MOS1
and MOS2) from the EPIC blank sky templates described by
\cite{Lumb02}. We have applied the same filtering procedure to the
background event files and extracted the spectra in the same region E1
and E2 in detector coordinates. Finally, the spectra have been
rebinned with the \texttt{grppha} task, so that there are at least 30
counts per energy bin.

\begin{table}
\centering
\caption[]{Count rates.}
\begin{tabular}{l c r}
\hline
Region  & counts/s & counts \\
\hline
E1 All MOS1  & $0.1096\pm0.0055$ & 1354 \\
E1 All MOS2  & $0.0899\pm0.0057$ & 1115 \\
\hline
E1 North MOS1& $0.0711\pm 0.0040$  & 878 \\
E1 North MOS2& $0.0654\pm 0.0041$  & 810 \\
\hline
E1 South MOS1& $0.0386 \pm 0.0038$ & 476 \\
E1 South MOS2& $0.0246 \pm 0.0039$ & 305 \\
\hline
E2 MOS1	& $0.010 \pm 0.001$ & 123 \\
E2 MOS2 & $0.007 \pm 0.001$ & 87 \\
\hline
\end{tabular}
\label{tbl:counts}
\end{table}

The count rates after background subtraction in all the regions
analyzed are given in Table~\ref{tbl:counts}.  Since the brightest
portion of the filament is at the northern edge of the elliptical
extraction region, the difference in the MOS1 and MOS2 count rates in
region E1 could arise from small uncertainties in the vignetting
corrections at the edge of the FOV of the XMM image centered on the
filament.

\begin{figure}[!htb]
\centering
\mbox{\psfig{figure=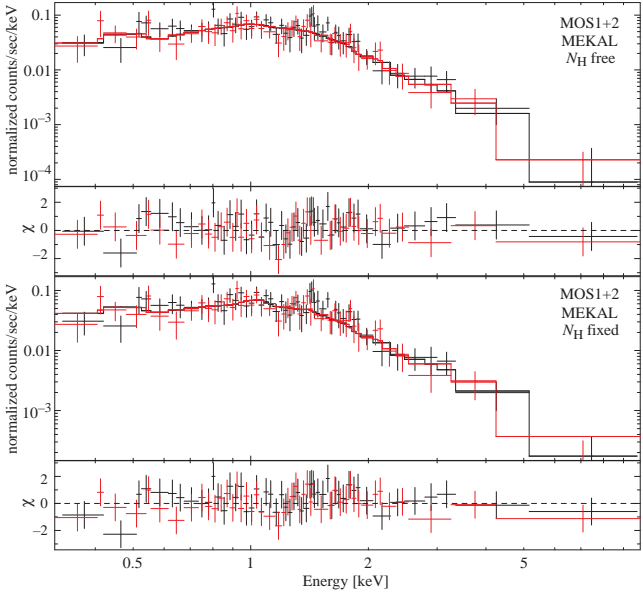,width=8.5cm}}
\caption[]{XMM-Newton MOS1 (black) and MOS2 (red) rebinned spectra of
the filament region (E1) with the best MEKAL fits superimposed. Top: fit
with the hydrogen column density free. Bottom: fit with $N_{\rm H}$ fixed
at the galactic value at the position of the filament.}
\label{fig:spectre}
\end{figure}

The spectral fits were done with XSPEC 11.2, with data in the range
[0.3--10~keV], simultaneously with MOS1 and MOS2. A MEKAL model
\citep{Kaastra,Liedahl} with photoelectric absorption given by
\cite{Balucinska} was used to fit the spectral data. Since the spectra
were rebinned, we have used standard $\chi^{2}$ minimization.
Fig.~\ref{fig:spectre} shows the MOS1 and MOS2 spectra, together with
the best MEKAL fits and residuals.

\begin{table}[!htb]
\centering
\caption[]{Spectral fits of the filament with the MEKAL model. Error bars are
90\% confidence limits.}
\begin{tabular}{l c c c c}
\hline
Region & $kT$ & $Z$  & $N_{\rm H}$ & $\chi^{2}/$dof\\
       & [keV] & [solar] & [$10^{20}\,$cm$^{-2}$] & \\
\hline
E1  & $1.9_{-0.4}^{+0.5}$ & $0.05_{-0.05}^{+0.13}$ & $8.2_{-3.6}^{+4.1}$ & 194/192\\
    & $2.4_{-0.3}^{+0.5}$ & $0.17_{-0.12}^{+0.17}$     &   3.16 (fixed)  & 200/193\\
\hline
E1 North & $2.0_{-0.4}^{+0.6}$ & $0.10_{-0.10}^{+0.19}$ & $6.2_{-3.6}^{+4.0}$ & 79/112\\
      & $2.4_{-0.4}^{+0.5}$ & $0.19_{-0.14}^{+0.19}$ &   3.16 (fixed)      & 81/113\\
\hline
E1 South & $1.5_{-0.5}^{+0.9}$ & $0.01_{-0.01}^{+0.21}$ & $14.0_{-8.3}^{+10.8}$ & 102/96\\
      & $2.5_{-0.7}^{+1.1}$ & $0.20_{-0.20}^{+0.53}$ &   3.16 (fixed)      & 107/97\\
\hline
\end{tabular}
\label{tbl:spectralfits}
\end{table}

Table~\ref{tbl:spectralfits} summarizes the spectral fits for the
filament. If $N_{\rm H}$ is left free to vary in the fit, it is not
well constrained, since its 90\% confidence interval is $(4.6 \leq
N_{\rm H} \leq 12.3) \times10^{20}\,$cm$^{-2}$, and the corresponding
X-ray temperature is in the range $1.5 \leq kT \leq 2.4$~keV. For the
metallicity, we only have an upper limit $Z \leq 0.18\ Z_{\odot}$,
which is almost the same as the mean metallicity obtained in the fit
with fixed $N_{\rm H}$.  The galactic neutral hydrogen column density
in the direction of the filament is $N_{\rm H}=3.16\times
10^{20}\,$cm$^{-2}$ \citep[]{Dickey}. When $N_{\rm H}$ is fixed to
this value, a MEKAL fit to the filament spectrum gives a gas
temperature of $2.0 \leq kT \leq 2.8\,$keV and a metallicity of
$0.04\leq Z \leq 0.33 Z_\odot$ (90\% confidence range). Due to the
small number of counts in region E2, it was not possible to derive
useful constraints on the spectral parameters.

Fluxes and luminosities were calculated with the best MEKAL fit and
are shown in Table~\ref{tbl:luminosityflux}.

\begin{table}[!htb]
\centering
\caption[]{Filament unabsorbed flux  and luminosity in two bands. Both 
cases, with $N_{\rm H}$ as a free and fixed parameter are shown. The
statistical uncertainties are about 10\%.}
\begin{tabular}{l l| c c | c c}
\hline
Region   & Band & \multicolumn{2}{c|}{(free $N_{\rm H}$)} & 
        \multicolumn{2}{c}{(fixed $N_{\rm H}$)}\\\relax
 & [keV]  & Flux  & Lumin. & Flux & Lumin.\\
\hline
E1 & 2.0--10.   &  3.0  &  4.5    & 3.5 & 5.2 \\
    & 0.5--4.5   &  8.4  &  11.9   & 7.7 & 10.8 \\
\hline
E1 North & 2.0--10.   & 2.5 &  3.7 & 2.8 & 4.1 \\
      & 0.5--4.5   & 6.4 &  9.0 & 6.1 & 8.5 \\
\hline
E1 South & 2.0--10.   & 0.8 &  1.1 & 1.0 & 1.5 \\
      & 0.5--4.5   & 2.6 &  3.7 & 2.1 & 3.0 \\
\hline
\end{tabular}
\label{tbl:luminosityflux}
\begin{flushleft}
    \vspace{-6pt}
    Note: Flux units are $10^{-13}$erg cm$^{-2}$ s$^{-1}$.\\
    Luminosity units are: $10^{42} h_{50}^{-2}\,$erg~s$^{-1}$
\end{flushleft}
\end{table}

We also attempted to fit a power-law to the spectra. Results are shown in
Table~\ref{tbl:powerlaw}. The resulting photon index is in the interval
2.5--3.0 and 1.8--2.0 when $N_{\rm H}$ is left free to vary or is fixed,
respectively.

\begin{table}[!htb]
\centering
\caption[]{Power-law fits of the filament spectra. Error bars are
90\% confidence limits.}
\begin{tabular}{c c c}
\hline
spectral index    &   $N_{\rm H}$ [$10^{20}\,$cm$^{-2}$] & $\chi^{2}/$dof\\
\hline
$2.7_{-0.3}^{+0.3}$ & $22.0_{-5.5}^{+6.1}$   &   201.0/193 \\
$1.9_{-0.1}^{+0.1}$ &   3.16 (fixed)         &   242.7/194 \\
\hline
\end{tabular}
\label{tbl:powerlaw}
\end{table}

The MEKAL and power law fits are of comparable statistical quality
when $N_{\rm H}$ is left free to vary; however, when $N_{\rm H}$ is
fixed, the MEKAL fit is significantly better (compare the bottom rows
of Tables 2 and 4). Since the fitted value of $N_{\rm H}$ for the
power law is unreasonably large and the value of $\chi^2$ unacceptable
when $N_{\rm H}$ is fixed at the Galactic value, we reject the power
law as a possible spectral model and favour thermal emission.

The computed fluxes and luminosities are actually from the filament
plus the cluster ICM at that position. The cluster count rate in
region E2 (after background subtraction) is given in
Table~\ref{tbl:counts}.  The cluster contribution to the fluxes and
luminosities presented in Table~\ref{tbl:luminosityflux} for the
filament (E1) is therefore about $\sim 10$\%.
 
Visual inspection of the filament (Fig.~\ref{fig:imminusmodel})
reveals a North $\rightarrow$ South gradient. We defined two halves of
the filament region (E1 north and south) and analysed them in the same
way as the whole E1 ellipse (as depicted in
Fig.~\ref{fig:imminusmodel}). The number counts in these two regions
are given in Table~\ref{tbl:counts}; they can be compared to the total
number of counts in the north and south halves of E2: 62 and 44
respectively. The MEKAL fitting results are summarized in
Table~\ref{tbl:spectralfits} and the fluxes and luminosities are given
in Table~\ref{tbl:luminosityflux}.  Within error bars, there is no
gradient of temperature or metal abundance.

\section{Discussion and conclusions}

Our XMM-Newton observations confirm that there really is a highly
elongated filamentary like structure extending from the South clump to
the south east of Abell~85 along the direction defined by all the
structures pointed out by Durret et al. (1998b).  The fact that the
spatial structure of the X-ray filament detected by XMM-Newton cannot
be exactly superimposed to that obtained from ROSAT data shows that it
is still difficult to determine exactly its structure. 

However, we have shown that the X-ray spectrum from this structure is
most likely thermal and its temperature is about 2.0 keV, consistent
with that of groups. This value is notably cooler than that of the
main cluster: the temperature map by Markevitch et al. (1998) shows
the presence of gas at about 3--4 keV in the region at a distance from
the cluster center at least as far as the northern part of the
ellipse. So, we appear to be seeing cool gas as it enters the cluster
core.

Another possibility is that the filament is associated with the wake
of cool stripped gas left behind by the south blob as it falls onto
the cluster. In this case, the ``filament'', whether it is diffuse or
made of groups, would not really be a filament in the large scale
structure formation sense. Besides X-ray observations with a much
better signal to noise ratio, which probably will have to wait for the
next generation of X-ray satellites, optical data can shed light on
this question. With this purpose, we intend to perform wide field
imaging in various bands to estimate galaxy photometric redshifts and
determine how galaxies are distributed in the ``filament'' area.

\begin{acknowledgements}

We acknowledge a demonstration of the XMM-Newton SAS software by
S\'ebastien Majerowicz and help from Sergio Dos Santos to install SAS
at the IAP and perform a preliminary data reduction. F.D. and
G.B.L.N. acknowledge financial support from the
USP/COFECUB. G.B.L.N. acknowledges support from FAPESP and
CNPq. W. Forman thanks the Max-Planck-Institute f\"ur Astrophysik for
its hospitality during the summer of 2002 and acknowledges support
from NASA Grant NAG5-10044. This work is based on observations
obtained with XMM-Newton, an ESA science mission with instruments and
contributions directly funded by ESA Member States and the USA (NASA).
May the referee be thanked for helping us improve the manuscript.

\end{acknowledgements}


\begin{thebibliography}{}

\bibitem[Balucinska-Church \& McCammon(1992)]{Balucinska} Balucinska-Church
M. \& McCammon D. 1992, ApJ 400, 699

\bibitem[Briel \& Henry (1995)]{Briel} Briel U. \& Henry P. 1995, A\&A 302, L9

\bibitem[Dickey \& Lockman(1990)]{Dickey} Dickey J.M. \& Lockman
F.J. 1990, Ann. Rev. Ast.  Astr. 28, 215

\bibitem[Durret et al.(1998a)]{Durret98op} Durret F., Felenbok P., Lobo C. \&
 Slezak E. 1998a, A\&A Suppl. 129, 281

\bibitem[Durret et al.(1998b)]{Durret98X} Durret F., Forman W., Gerbal D., 
Jones C. \& Vikhlinin A. 1998b, A\&A 335, 41

\bibitem[Frenk, White \& Davis(1983)]{Frenk83} Frenk C.S., White S.D.M. \&
Davis M. 1983, ApJ 271, 417

\bibitem[Jansen et al.(2001)]{Jansen01} Jansen F., Lumb D., Altieri B. et 
al. 2001, A\&A 365, L1

\bibitem[Jenkins et al.(1998)]{Jenkins98} Jenkins A., Frenk C.S., Pearce F.R.
et al. 1998, {ApJ 499, 20}

\bibitem[Kaastra \& Mewe(1993)]{Kaastra} Kaastra J.S. \& Mewe R. 1993, 
A\&AS 97, 443 

\bibitem[Katgert et al. (1996)]{Katgert}
Katgert P., Mazure A., Perea J. et al. 1996, A\&A 310, 8

\bibitem[Kempner, Sarazin \& Ricker (2002)]{Kempner} Kempner J.C.,
Sarazin C.L. \& Ricker P.M. 2002, ApJ 579, 236

\bibitem[Liedahl et al.(1995)]{Liedahl} Liedahl D.A., Osterheld A.L. \&
Goldstein W.H. 1995, ApJ 438, L115 

\bibitem[Lumb et al.(2002)]{Lumb02} Lumb D.H., Warwick R.S., Page M. \& 
De Luca A. 2002, A\&A 389, 93

\bibitem[Markevitch et al. (1998)]{Markevitch}
Markevitch M., Forman W.R., Sarazin C.L., Vikhlinin A. 1998 ApJ, 503, 77 

\bibitem[Pierre et al.(2000)]{Pierre} Pierre M., Bryan G. \& Gastaud R. 2000, 
A\&A 356, 403

\bibitem[Tittley \& Henriksen(2001)]{Tittley} Tittley E.R. \& Henriksen M.
2001, ApJ 563, 673

\bibitem[Schmalzing \& Diaferio(2000)]{Schmalzing} Schmalzing J. \& 
Diaferio A. 2000, MNRAS 312, 638

\bibitem[Vikhlinin, Forman \& Jones (1997)]{Vikhlinin}
Vikhlinin A., Forman W. \& Jones C. 1997, ApJL 474, L7

\bibitem[Wang, Connolly \& Brunner (1997)]{Wang}
Wang Q.D., Connolly A. \& Brunner R. 1997, ApJ 487, L13

\bibitem[West, Jones \& Forman (1995)]{West95}
West M.J., Jones C. \& Forman W. 1995, ApJL 451, L5

\end{thebibliography}
\end{document}